\documentclass{article}
\usepackage{ijcai26}

\usepackage{times}
\usepackage{soul}
\usepackage{url}
\usepackage{enumitem}
\usepackage[hidelinks]{hyperref}
\usepackage[utf8]{inputenc}
\usepackage[small]{caption}
\usepackage{graphicx}
\usepackage{amsmath}
\usepackage{amsthm}
\usepackage{amssymb}
\usepackage{booktabs}
\usepackage{algorithm}
\usepackage{algorithmic}
\usepackage{textgreek}
\usepackage{tikz}
\usepackage{xcolor}
\usepackage{tabularx}
\usetikzlibrary{arrows.meta, shapes.geometric, positioning, calc, backgrounds, fit}

\usepackage[switch]{lineno}
\usepackage{booktabs}
\usepackage{bm}

\title{C2NP: A Benchmark for Learning Scale-Dependent\\ Geometric Invariances in 3D Materials Generation}

\author{
\vspace{1cm}
\normalfont\small
\begin{tabular}{c c}
\begin{minipage}[t]{0.45\linewidth}\centering
Can Polat\\
Texas A\&M University\\
{\footnotesize \texttt{can.polat@tamu.edu}}
\end{minipage}
&
\begin{minipage}[t]{0.45\linewidth}\centering
Erchin Serpedin\\
Texas A\&M University\\
{\footnotesize \texttt{eserpedin@tamu.edu}}
\end{minipage}
\\[12mm]
\begin{minipage}[t]{0.45\linewidth}\centering
Mustafa Kurban\textsuperscript{*}\\
Ankara University\\
Texas A\&M University at Qatar\\
{\footnotesize \texttt{kurbanm@ankara.edu.tr}}
\end{minipage}
&
\begin{minipage}[t]{0.45\linewidth}\centering
Hasan Kurban\textsuperscript{*}\\
Hamad Bin Khalifa University\\
{\footnotesize \texttt{hkurban@hbku.edu.qa}}
\end{minipage}
\end{tabular}
\thanks{Corresponding authors.}
}

\begin{document}

\maketitle

\begin{abstract}
Generative models for materials have achieved strong performance on periodic bulk crystals, yet their ability to generalize across scale transitions to finite nanostructures remains largely untested. We introduce \emph{Crystal-to-Nanoparticle (C2NP)}, a systematic benchmark for evaluating generative models when moving between infinite crystalline unit cells and finite nanoparticles, where surface effects and size-dependent distortions dominate. C2NP defines two complementary tasks: (i) generating nanoparticles of specified radii from periodic unit cells, testing whether models capture surface truncation and geometric constraints; and (ii) recovering bulk lattice parameters and space-group symmetry from finite particle configurations, assessing whether models can infer underlying crystallographic order despite surface perturbations. Using diverse materials as a structurally consistent testbed, we construct over 170{,}000 nanoparticle configurations by carving particles from supercells derived from DFT-relaxed crystal unit cells, and introduce size-based splits that separate interpolation from extrapolation regimes. Experiments with state-of-the-art approaches, including diffusion, flow-matching, and variational models, show that even when losses are low, models often fail geometrically under distribution shift, yielding large lattice-recovery errors and near-zero joint accuracy on structure and symmetry. Overall, our results suggest that current methods rely on template memorization rather than scalable physical generalization. C2NP offers a controlled, reproducible framework for diagnosing these failures, with immediate applications to nanoparticle catalyst design, nanostructured hydrides for hydrogen storage, and materials discovery. Dataset and code are available at \url{https://github.com/KurbanIntelligenceLab/C2NP}.
\end{abstract}

\begin{figure*}[t]
\centering
\resizebox{1\linewidth}{!}{%
\begin{tikzpicture}[
  font=\sffamily\footnotesize,
  >=Stealth,
  base_box/.style={
    draw=black!75,
    fill=white,
    line width=0.7pt,
    rounded corners=2pt,
    inner sep=4pt,
    minimum height=1.2cm
  },
  section_title/.style={
    fill=black!5,
    draw=black!75,
    font=\bfseries\footnotesize,
    inner sep=4pt
  },
  accent_bar/.style={ line width=1.6pt, draw=#1 },
  arrow_style/.style={
    ->, draw=black!70, line width=0.9pt,
    shorten >=2pt, shorten <=2pt
  }
]

\definecolor{oi_blue}{RGB}{0,114,178}
\definecolor{oi_orange}{RGB}{230,159,0}
\definecolor{oi_green}{RGB}{0,158,115}
\definecolor{soft_gray}{RGB}{120,120,120}

\def\vpad{38pt}        
\def\titlegap{\vpad}  
\def\rowsep{\vpad+10}    

\coordinate (C1) at (3.7,0);
\coordinate (C2) at (12.3,0);
\coordinate (C3) at (20.9,0);

\node[section_title, anchor=west,
      fill=oi_green!20, draw=oi_green!80!black] (S1T) at (0,0)
  {I.\;Structural Construction};

\node[base_box, minimum width=7.2cm,
      draw=oi_green!80!black, fill=oi_green!8] (UC)
  at ($(C1)+(0,-\titlegap)$) {
  \begin{tabular}{c}
    \textbf{Unit cell $\mathcal{U}$} \\
    $\Lambda = (a,b,c,\alpha,\beta,\gamma)$
  \end{tabular}
};

\node[base_box, minimum width=7.2cm,
      draw=oi_green!70!black, fill=oi_green!12] (SC)
  at ($(C2)+(0,-\titlegap)$) {
  \begin{tabular}{c}
    \textbf{Supercell} \\
    Tiling $\mathcal{T}=20^3$
  \end{tabular}
};

\node[base_box, minimum width=7.2cm,
      draw=oi_green!60!black, fill=oi_green!16] (NP)
  at ($(C3)+(0,-\titlegap)$) {
  \begin{tabular}{c}
    \textbf{Nanoparticle $\mathcal{P}_R$} \\
    Carve sphere of radius $R$ around $\mathbf{x}_0$
  \end{tabular}
};

\draw[arrow_style, draw=soft_gray!90] (UC.east) -- (SC.west);
\draw[arrow_style, draw=soft_gray!90] (SC.east) -- (NP.west);

\draw[accent_bar=oi_green!90!black] (UC.north west) -- (UC.north east);
\draw[accent_bar=oi_green!90!black] (SC.north west) -- (SC.north east);
\draw[accent_bar=oi_green!90!black] (NP.north west) -- (NP.north east);

\node[section_title, anchor=west,
      fill=oi_blue!20, draw=oi_blue!80!black] (S2T)
  at ($(UC.south west)!0.5!(NP.south east) + (-12.2, -\rowsep)$) {};
\node[section_title, anchor=west,
      fill=oi_blue!20, draw=oi_blue!80!black] (S2T)
  at ($(S1T.west |- UC.south) + (0,-\rowsep)$)
  {II.\;Orientational Augmentation Over $SO(3)$};

\node[base_box, minimum width=7.2cm,
      draw=oi_blue!80!black, fill=oi_blue!8] (AUG_TRAIN)
  at ($(C1 |- S2T)+(0,-\titlegap)$) {
  \begin{tabular}{l}
    \textbf{Training set $\mathcal{Q}_{\text{train}}$} \\
    Sparse grid: $\theta_{\text{train}}=15^\circ$ ($N\!\approx\!59$) \\
    Deterministic seeding over $SO(3)$
  \end{tabular}
};
\draw[accent_bar=oi_blue!90!black] (AUG_TRAIN.north west) -- (AUG_TRAIN.north east);

\node[base_box, minimum width=7.2cm,
      draw=oi_blue!85!black, fill=oi_blue!14] (AUG_ID)
  at ($(C2 |- S2T)+(0,-\titlegap)$) {
  \begin{tabular}{l}
    \textbf{ID set $\mathcal{Q}_{\text{ID}}$} \\
    Medium grid: $\theta_{\text{ID}}=12^\circ$ ($N\!\approx\!92$) \\
    Split margin: $d\ge\delta_{\text{ID}}=6^\circ$, offset $R_{\text{ID}}q$
  \end{tabular}
};
\draw[accent_bar=oi_blue!95!black] (AUG_ID.north west) -- (AUG_ID.north east);

\node[base_box, minimum width=7.2cm,
      draw=oi_blue!90!black, fill=oi_blue!20] (AUG_OOD)
  at ($(C3 |- S2T)+(0,-\titlegap)$) {
  \begin{tabular}{l}
    \textbf{OOD set $\mathcal{Q}_{\text{OOD}}$} \\
    Dense grid: $\theta_{\text{OOD}}=9^\circ$ ($N\!\approx\!163$) \\
    Split margin: $d\ge\delta_{\text{OOD}}=4.5^\circ$, offset $R_{\text{OOD}}q$
  \end{tabular}
};
\draw[accent_bar=oi_blue!100!black] (AUG_OOD.north west) -- (AUG_OOD.north east);

\draw[draw=soft_gray!90, line width=0.9pt] (AUG_TRAIN.east) -- (AUG_ID.west);
\draw[draw=soft_gray!90, line width=0.9pt] (AUG_ID.east) -- (AUG_OOD.west);

\node[section_title, anchor=west,
      fill=oi_orange!20, draw=oi_orange!80!black] (S3T)
  at ($(S2T.west |- AUG_TRAIN.south) + (0,-\rowsep)$)
  {III.\;Benchmark Tasks};

\node[base_box, minimum width=11.6cm,
      draw=oi_orange!80!black, fill=oi_orange!15] (T1)
  at ($(6.0,0 |- S3T)+(0,-\titlegap)$) {
  \begin{tabular}{l}
    \textbf{Task 1: Unit cell $\rightarrow$ nanoparticle} \\
    Input: $(\mathcal{U},R)$, Output: $\mathcal{P}_R$ \\
    Metrics: RMSD, Hausdorff distance, hull volume...
  \end{tabular}
};
\draw[accent_bar=oi_orange!90!black] (T1.north west) -- (T1.north east);

\node[base_box, minimum width=11.6cm,
      draw=oi_orange!90!black, fill=oi_orange!25] (T2)
  at ($(18.8,0 |- S3T)+(0,-\titlegap)$) {
  \begin{tabular}{l}
    \textbf{Task 2: Nanoparticle $\rightarrow$ lattice} \\
    Input: $\mathcal{P}_R$, Output: $(\Lambda,\Gamma)$ \\
    Metrics: lattice RMSE, space-group accuracy, joint recovery
  \end{tabular}
};
\draw[accent_bar=oi_orange!100!black] (T2.north west) -- (T2.north east);

\draw[draw=soft_gray!90, line width=0.9pt] (T1.east) -- (T2.west);

\def\rightoffset{4mm}

\draw[->, draw=soft_gray!90, line width=0.9pt]
  (NP.east) -- ($ (NP.east) + (\rightoffset,0) $);

\draw[draw=soft_gray!90, line width=0.9pt]
  ($ (NP.east) + (\rightoffset,0) $) --
  ($ (AUG_OOD.east) + (\rightoffset,0) $) --
  ($ (T2.east) + (\rightoffset,0) $);

\draw[->, draw=soft_gray!90, line width=0.9pt]
  ($ (AUG_OOD.east) + (\rightoffset,0) $) -- (AUG_OOD.east);

\draw[->, draw=soft_gray!90, line width=0.9pt]
  ($ (T2.east) + (\rightoffset,0) $) -- (T2.east);

\def\leftoffset{-6mm} 

\draw[->, draw=soft_gray!90, line width=0.9pt] 
  (UC.west) -- ($ (UC.west) + (\leftoffset,0) $);

\draw[draw=soft_gray!90, line width=0.9pt]
  ($ (UC.west) + (\leftoffset,0) $) -- 
  ($ (AUG_TRAIN.west) + (\leftoffset,0) $) -- 
  ($ (T1.west) + (\leftoffset,0) $);

\draw[->, draw=soft_gray!90, line width=0.9pt] 
  ($ (AUG_TRAIN.west) + (\leftoffset,0) $) -- (AUG_TRAIN.west);

\draw[->, draw=soft_gray!90, line width=0.9pt] 
  ($ (T1.west) + (\leftoffset,0) $) -- (T1.west);
  
\begin{scope}[on background layer]
  \node[
    fit=(S1T)(UC)(SC)(NP)(S2T)(AUG_TRAIN)(AUG_ID)(AUG_OOD)(S3T)(T1)(T2),
    fill=black!2,
    draw=none,
    rounded corners=2pt,
    inner xsep=4pt,
    inner ysep=\vpad -15
  ] {};
\end{scope}

\end{tikzpicture}%
}
\caption{C2NP data generation and evaluation pipeline. The first stage (green) constructs finite nanoparticles from periodic crystallographic inputs: starting from each DFT-optimized unit cell \(\mathcal{U}\) with lattice parameters \(\Lambda = (a,b,c,\alpha,\beta,\gamma)\), a \(20\times20\times20\) supercell \(\mathcal{T}\) is built to ensure sufficient spatial extent, after which a spherical cluster \(\mathcal{P}_R\) is carved by retaining atoms within radius \(R\) of a chosen center \(\mathbf{x}_0\). The second stage (blue) applies rotational augmentation over \(\mathrm{SO}(3)\) using a geodesic separation constraint on unit quaternions to generate three stratified orientation grids: a sparse training set \(\mathcal{Q}_{\mathrm{train}}\) with threshold \(\theta_{\mathrm{train}}=15^\circ\), an interpolative in-distribution set \(\mathcal{Q}_{\mathrm{ID}}\) with \(\theta_{\mathrm{ID}}=12^\circ\) and exclusion margin \(\delta_{\mathrm{ID}}=6^\circ\), and a dense extrapolative out-of-distribution set \(\mathcal{Q}_{\mathrm{OOD}}\) with \(\theta_{\mathrm{OOD}}=9^\circ\) and margin \(\delta_{\mathrm{OOD}}=4.5^\circ\). Deterministic quaternion seeding and fixed offset rotations \(R_{\mathrm{ID}}\) and \(R_{\mathrm{OOD}}\) enforce non-overlapping geodesic neighborhoods between train, ID, and OOD orientations while decorrelating orientation manifolds. The third stage (orange) defines two complementary benchmark tasks: a forward problem mapping unit-cell parameters and target radius \((\mathcal{U},R)\) to the corresponding nanoparticle geometry \(\mathcal{P}_R\), and an inverse problem recovering lattice parameters and symmetry \((\Lambda,\Gamma)\) from a nanoparticle \(\mathcal{P}_R\), providing a unified testbed for both generative and inverse lattice--nanoparticle modeling.}
\label{fig:nanoscale_pipeline}
\end{figure*}
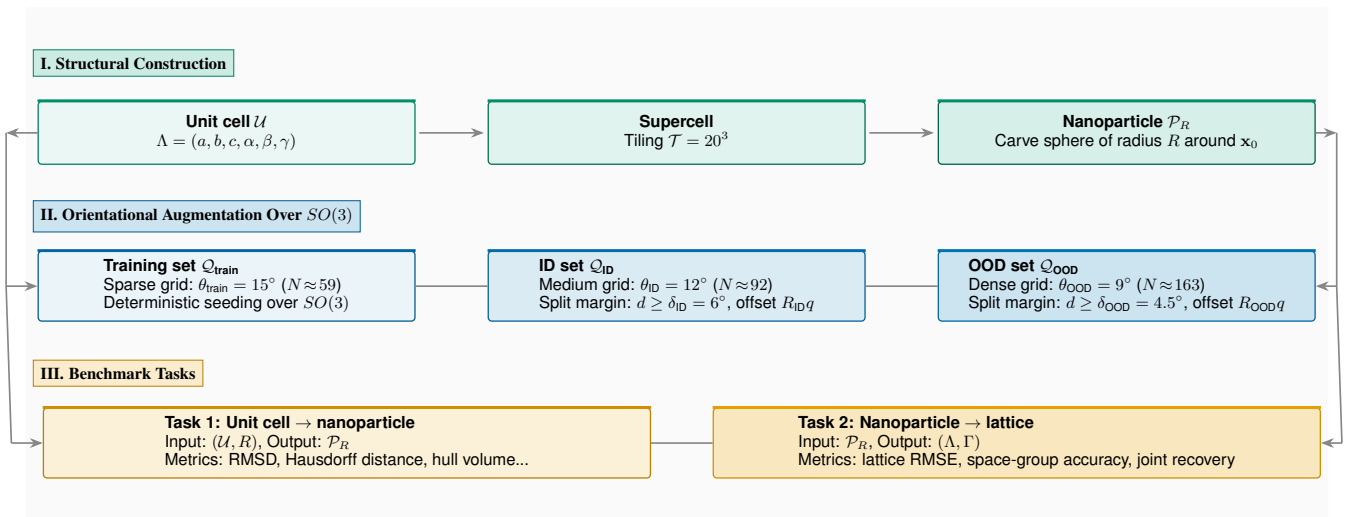

\begin{figure*}[h]
\centering
\includegraphics[width=0.98\textwidth]{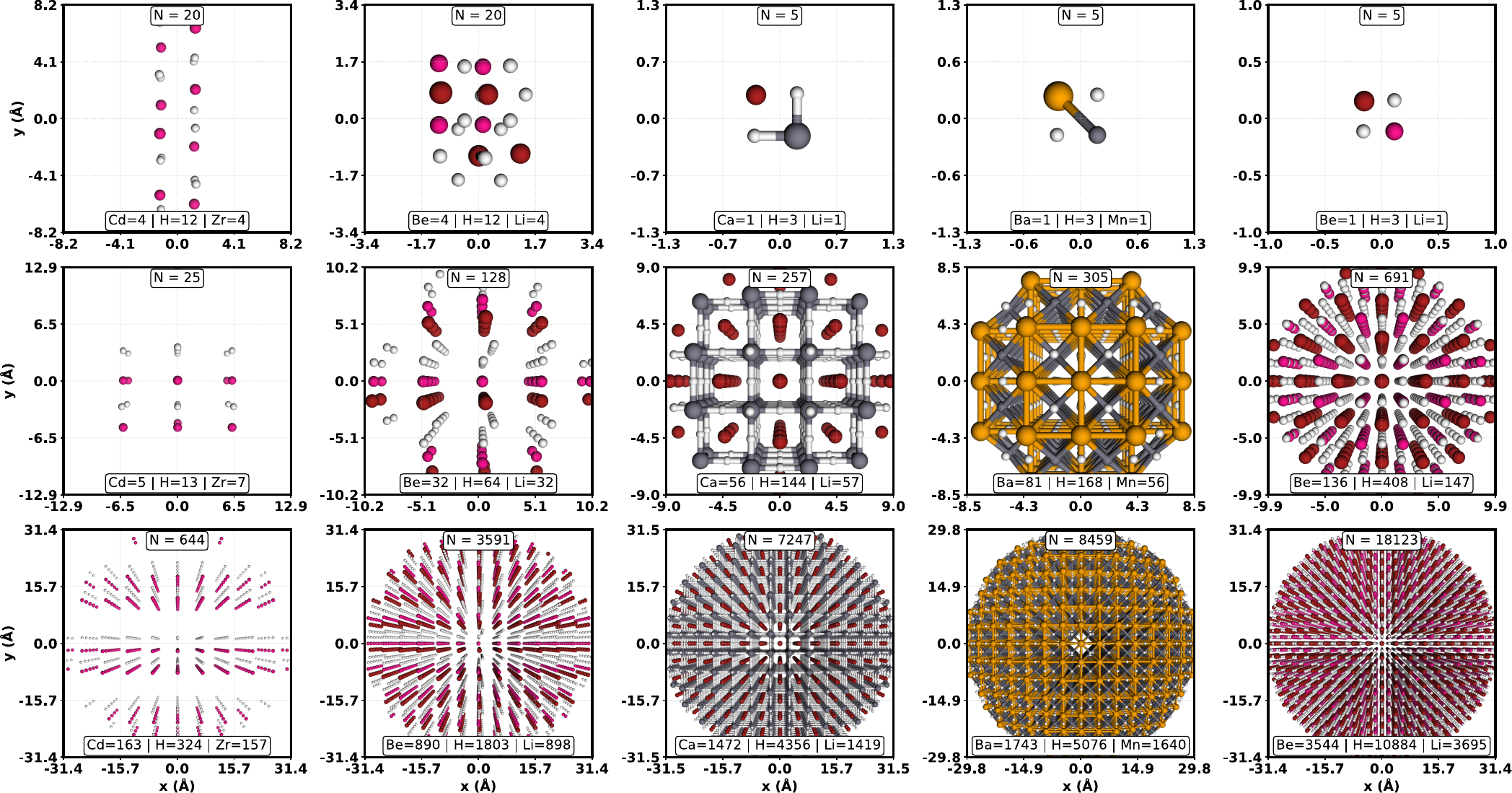}
\caption{Representative crystal-to-nanoparticle transformations in the C2NP dataset, derived from DFT-validated crystallographic reference structures. Columns (left to right) correspond to ZrCdH$_3$, LiBeH$_3$, LiCaH$_3$, BaMnH$_3$, and LiBeH$_3$. Rows show the reference unit cell (top), a nanoparticle carved at $R=10$,\AA\ (middle), and a nanoparticle carved at $R=30$,\AA\ (bottom). Each panel reports the total atom count ($N$) and elemental composition. Across the dataset, particle sizes span more than four orders of magnitude, from a minimum of $N=4$ atoms for the smallest particles to a maximum of $N=18{,}123$ atoms for the largest configurations. Atom colors follow the standard CPK scheme: H (white), Li (purple), Be (dark green), Ca (green), Ba (light green), Mn (purple), Zr (dark gray), and Cd (light gray).}
\label{fig:materials_showcase}
\end{figure*}
\section{Introduction}\label{sec:intro}

Computational materials modeling generally falls into two distinct but interconnected domains. The first domain concerns infinite periodic crystals, characterized by unit cells that encode lattice constants, atomic basis vectors, and space group operations \cite{tarantino2017structural}. This crystallographic representation, standardized through CIF formats, distills the translational symmetry of bulk solids and forms the conceptual backbone of solid-state physics and quantum chemistry \cite{kittel2018introduction}. The second domain encompasses nanoparticles, finite atomic assemblies where translational periodicity is absent. Here, surface facets, edge sites, and reduced coordination numbers give rise to structural relaxations, electronic localization, and quantum size effects that profoundly modify functional properties \cite{pizzagalli2001structure}. Many technologically relevant materials exist at the intersection of these descriptions, and characterizing how bulk order transitions to finite-scale behavior is essential for understanding catalysis, optoelectronics, thermodynamic stability, and charge transport \cite{vergara2017microed}.

Although recent progress in machine learning has transformed materials discovery, connecting bulk and nanoscale regimes remains an open problem \cite{li2023critical}. Neural models trained on crystallographic databases often generalize well within periodic systems but exhibit significant degradation when tasked with nanoparticle generation or lattice inference from finite configurations \cite{gleason2024random}. Common failure modes include incorrect space group assignment, erroneous lattice constant predictions, and poor representation of size-driven surface effects \cite{jain2024machine}. Current benchmarks expose these shortcomings: CSPBench demonstrates that leading crystal prediction methods collapse when evaluation distributions deviate from training data \cite{wei2024cspbench}. Meanwhile, application-focused perovskite datasets prioritize property prediction tasks, such as bandgap estimation or solar cell efficiency, without addressing bidirectional structural transformations or controlled size-dependent variation \cite{kusaba2022crystal}.

To address this gap, we present \textit{C2NP}, a benchmark that systematically tests whether generative models can handle the structural transformations inherent to bulk-to-nanoparticle transitions. Our approach leverages perovskite hydrides, a material family with well-established relevance to energy storage and heterogeneous catalysis \cite{kuo2024overview}, as a structurally coherent platform. We construct a dataset pairing DFT-validated unit cells with  about 172,000 nanoparticle configurations derived by spherical carving across radii from 6 to 30~\AA\ with 1 \AA\ steps. This construction strategy isolates the impact of finite size and surface effects while maintaining chemical diversity, avoiding the confounding influence of disparate crystal topologies that would complicate interpretation \cite{csopu2011finite}. The benchmark is organized around two complementary evaluation protocols: forward generation, where models must produce nanoparticles of specified radius from a given unit cell (requiring understanding of how periodic lattices terminate at finite boundaries), and inverse reconstruction, where models infer bulk lattice parameters and space group symmetry from nanoparticle structures despite surface disorder and coordinate ambiguity. This bidirectional design probes whether learned representations encode scalable structural principles or merely memorize correlations within the training distribution. C2NP thus provides not only a diagnostic benchmark for identifying generalization failures but also a foundation for developing architectures that can reason about physical scale in crystalline matter. Figure~\ref{fig:nanoscale_pipeline} outlines the C2NP benchmark pipeline, including dataset construction and evaluation flow, and Figure~\ref{fig:materials_showcase} visualizes representative crystal-to-nanoparticle transformations derived from DFT-validated unit cells.

\section{Related Work}\label{sec:relatedWork}

\subsection{Bulk Crystals and Finite Nanostructures}

The unit cell, comprising lattice vectors, space group operations, and atomic coordinates, serves as the fundamental descriptor for crystalline solids \cite{anosova2024importance}. This representation provides the basis for crystallographic databases, electronic structure calculations, and most computational materials repositories \cite{hellenbrandt2004inorganic}. By encoding infinite periodicity in a minimal repeating unit, the unit cell offers an exceptionally compact and expressive structural blueprint \cite{jain2013commentary}. Real functional materials, however, often exist as finite nanoparticles rather than extended crystals \cite{baig2021nanomaterials}. At these reduced length scales, surfaces constitute a dominant fraction of the structure, and the absence of translational periodicity leads to coordination deficits, facet-dependent reconstructions, and electronic confinement phenomena that reshape material behavior \cite{cheng2024perovskite}. Surface atoms, edge sites, and boundary terminations introduce structural and electronic perturbations absent in bulk descriptions \cite{zhang2023surface}. Bridging unit cell and nanoparticle representations is therefore critical across application domains \cite{cheng2024perovskite}. Catalytic activity, for instance, arises largely from surface site geometry and defect density, while size-tunable photoluminescence in quantum dots stems from quantum confinement within nanoscale boundaries \cite{ye2024strongly}. Nonetheless, generating nanoparticles from crystallographic input or extracting bulk lattice parameters from finite configurations remains challenging due to the non-additive effects of surfaces and the geometric complexity introduced by symmetry breaking \cite{zhang2023surface}.

\subsection{Existing Benchmarks and Structural Databases}

Progress in computational materials discovery has been propelled by large-scale structure repositories. The Materials Project \cite{jain2013commentary} and OQMD \cite{kirklin2015open} curate significant amount of periodic crystal structures, enabling property prediction and high-throughput screening. CSPBench provides standardized tasks for crystal structure prediction, emphasizing thermodynamic ranking and bulk phase identification. PubChemQC \cite{kim2025pubchem} supplies quantum-chemical data for molecular systems at scale, facilitating transfer learning between domains. The Perov-5 dataset \cite{perov5a,perov5b} aggregates perovskite crystal data with computed properties, while OC20 \cite{chanussot2021open} and OC22 \cite{tran2023open} benchmark adsorbate interactions on metal surfaces with force annotations. CrysMTM \cite{polat2025crysmtm} introduces prediction under thermal effects for crystal graph networks. Despite their impact, these resources universally focus on bulk or surface-slab configurations and do not systematically incorporate finite nanoparticle structures or size-dependent variations. Complementary benchmarks address model robustness and transfer learning. Matbench \cite{dunn2020benchmarking} assembles property prediction tasks across diverse datasets to assess supervised learning pipelines. Within the perovskite literature, datasets targeting band-gap regression, photovoltaic performance, and phase stability have emerged to support application-specific objectives \cite{talapatra2023band}. While valuable for targeted materials design, these collections do not interrogate the structural mapping between infinite lattices and finite clusters, nor do they explore systematic radius variation \cite{li2018predicting}. No prior benchmark pairs unit cells with radius-resolved nanoparticle ensembles, precluding rigorous evaluation of scale-transfer capabilities. This absence limits our understanding of whether models learn geometric principles that generalize across length scales or merely fit empirical correlations within narrow training regimes. Addressing this limitation motivates the construction of C2NP.

\subsection{Generative Architectures for Crystal Modeling}

Machine learning for crystalline materials has evolved rapidly alongside the growth of structural databases. Initial efforts employed graph convolutional networks \cite{cgcnn}, and message-passing frameworks \cite{klipfel2023equivariant} for property prediction. Subsequent architectures incorporating equivariance constraints demonstrated significant improvements in force field accuracy by respecting rotational and translational symmetries \cite{nequip,liao2022equiformer,kurban2026multimodal}. Generative modeling of crystal structures has likewise advanced through variational autoencoders \cite{cvae_crystals}, denoising diffusion frameworks \cite{khastagir2025crysldm}, flow-based methods conditioned on large language models \cite{sriram2024flowllm}, and autoregressive generation over graph topologies \cite{antunes2024crystal}. However, the absence of datasets linking periodic unit cells to size-resolved nanoparticles has prevented systematic assessment of how these models perform across scale transitions. Current benchmarks cannot determine whether learned representations capture the geometric logic governing finite-size effects or whether they simply interpolate within the distribution of bulk training data \cite{butler2018machine}. C2NP fills this evaluation gap by providing explicitly paired structures and bidirectional tasks that test both nanoparticle synthesis from lattice blueprints and lattice inference from finite atomic clusters.

\section{Benchmark Construction}\label{sec:method}

\subsection{Generating Nanoparticles from Unit Cells}
\label{sec:nano_construction}

We assembled a collection of compositions with documented relevance to catalysis and energy storage, extracting their DFT-optimized lattice parameters from published literature. Starting from each crystallographic information file, we constructed the primitive unit cell and subsequently expanded it into a supercell of sufficient extent to accommodate the desired nanoparticle radii. Finite clusters were generated by applying spherical truncation: atoms whose coordinates lie within a sphere of radius $R$ centered at $\mathbf{x}_0$ are retained, while those beyond this boundary are discarded. We deliberately refrain from further structural relaxation or surface passivation, isolating purely geometric size effects while maintaining consistency with validated bulk lattice parameters. The whole process  is  highlighted in Figure \ref{fig:nanoscale_pipeline} under I. Structural Construction.

\noindent\textbf{Supercell dimensions.}
Each composition is replicated into a $20\times20\times20$ supercell by tiling the primitive cell along the $a$, $b$, and $c$ axes, producing linear dimensions $L_i=20\,a_i$ for $i\in\{a,b,c\}$. This ensures $L_i \ge 2R_{\max}+\Delta$ for the largest carved radius $R_{\max}=30\,\text{\AA}$ with a buffer $\Delta\approx 5$–$10\,\text{\AA}$, preventing interactions between the nanoparticle and its periodic images.

\noindent\textbf{Spherical truncation protocol.}
A nanoparticle of specified radius $R$ is extracted by selecting all atoms satisfying the spatial criterion
\[
\mathcal{C}_R=\{\,\mathbf{x}_i \in \mathbf{R}^3 \mid \|\mathbf{x}_i-\mathbf{x}_0\|\le R\,\},
\]
where $R \in \{6,7,8,\dots,30\}\,\text{\AA}$. This procedure produces a systematic series of finite structures per composition, spanning regimes where surface termination dominates (small $R$) to those approaching bulk-like coordination (large $R$).

\subsection{Orientational Augmentation and Data Partitioning}

To eliminate directional bias and ensure rigorous train-test separation, we augment each nanoparticle configuration through rotational sampling over $\mathrm{SO}(3)$ \cite{shoemake1985animating}. Rotations are parameterized by unit quaternions $q\in\mathbf{H}$ with $\|q\|=1$. The geodesic separation between rotations $q_i$ and $q_j$ is quantified as
\[
d(q_i,q_j)=2\arccos\!\left(\big|\langle q_i,q_j\rangle\big|\right),
\]
where $\langle\cdot,\cdot\rangle$ denotes the standard inner product in $\mathbf{R}^4$. We employ a greedy algorithm to construct a rotation set $\mathcal{Q}(\theta)$ satisfying
\[
d(q_i,q_j)\ge \theta \quad \forall\, i\neq j,
\]
with the angular threshold $\theta$ governing sampling density. Approximating rotational coverage via spherical caps yields the asymptotic estimate
\[
N(\theta)\;\approx\;\frac{2}{1-\cos\theta},
\]
indicating that tighter angular constraints increase sample count \cite{kuffner2004effective}. For the thresholds adopted here,
\[
N(15^\circ)\approx 59,\qquad N(12^\circ)\approx 92,\qquad N(9^\circ)\approx 163.
\]

Denote the deterministically seeded training rotation set as $\mathcal{Q}_{\mathrm{train}}$. To enforce disjoint evaluation sets, candidate test quaternions are admitted only if they maintain a minimum angular separation from all training rotations,
\[
d(q,q')\;\ge\;\delta_{\mathrm{split}}\quad \forall\, q'\in\mathcal{Q}_{\mathrm{train}},
\]
where $\delta_{\mathrm{ID}}=6^\circ$ and $\delta_{\mathrm{OOD}}=4.5^\circ$. Exploiting the geodesic formula, this constraint simplifies to
\[
|\langle q,q'\rangle|\;\le\;\cos\!\big(\tfrac{1}{2}\delta_{\mathrm{split}}\big).
\]
Additionally, fixed rotation offsets $R_{\mathrm{ID}}$ and $R_{\mathrm{OOD}}$ are applied via left-multiplication,
\[
q_{\mathrm{eff}}=R_{\mathrm{split}}\cdot q,
\]
using deterministic Euler angles (e.g., $(6^\circ, 8^\circ, 12^\circ)$ for ID, $(15^\circ, 25^\circ, 35^\circ)$ for OOD) to further decorrelate orientation distributions while preserving geodesic separations.

Size-based partitioning assigns:
\begin{eqnarray*}
\mathcal{R}_{\mathrm{ID}}&=&\{10,11,17,21,24,26\}, \\
\mathcal{R}_{\mathrm{OOD}}&=&\{6,7,29,30\},
\end{eqnarray*}
with all intermediate radii reserved for training. Rotational sampling densities are stratified: $\theta_{\mathrm{train}}=15^\circ$ for training, $\theta_{\mathrm{ID}}=12^\circ$ for interpolation testing, and $\theta_{\mathrm{OOD}}=9^\circ$ for extrapolation evaluation.

\subsection{Rationale for Partitioning Strategy}

The radius and angular spacing choices implement a coarse-to-dense sampling strategy balancing computational efficiency, statistical reliability, and evaluation rigor. Training data spans the central radius range using a sparse orientation grid ($\theta_{\mathrm{train}}=15^\circ$), providing broad coverage without excessive redundancy. In-distribution testing targets mid-range radii ($10,11,17,21,24,26$) with moderately denser sampling ($\theta_{\mathrm{ID}}=12^\circ$, yielding approximately $92$ orientations per structure) to stabilize metric estimates without introducing novel geometric regimes. Out-of-distribution evaluation focuses on size extremes ($6,7,29,30$) where surface-to-volume ratios ($S/V\!\sim\!3/R$) amplify finite-size effects, employing the densest sampling ($\theta_{\mathrm{OOD}}=9^\circ$, approximately $163$ orientations) to minimize variance in error estimation. All evaluation quaternions maintain strict geodesic separation from training via the exclusion margins $\delta_{\mathrm{ID}}$ and $\delta_{\mathrm{OOD}}$, precluding any distributional overlap. Formally, orientation-averaged error for a given radius is computed as
\[
\overline{E}(R)=\frac{1}{|\mathcal{Q}_R|}\sum_{q\in \mathcal{Q}_R} E\!\left(\mathcal{R}_q(\text{prediction}),\;\mathcal{R}_q(\text{reference})\right),
\]
where $\mathcal{R}_q$ denotes the rotation operator. Denser orientation grids reduce estimator variance, particularly critical for OOD assessment where predictions are most variable. Structures exhibiting high crystallographic symmetry undergo deduplication, retaining only distinct rotational equivalents. Complete partitioning specifications are provided in Supplementary~\ref{sec:appenSplits}.

\subsection{Compositional Characteristics}

The A-sublattice comprises ten species, predominantly alkali and alkaline earth metals, with the most frequent being Li (21 instances), Na (16), and K (14). Electronegativity values for this sublattice exhibit mean $0.999 \pm 0.218$, while ionic radii span $1.160 \pm 0.300$\,\AA, reflecting substantial size heterogeneity that influences structural distortion and thermodynamic stability. The B-sublattice displays greater chemical diversity, encompassing 27 elements concentrated in transition metal and metalloid groups, with V and Rh appearing most frequently (8 instances each), followed by Cu and Zn (6 each). In contrast to the A-site, B-site properties cluster more tightly: electronegativity averages $1.543 \pm 0.114$ and ionic radii $0.694 \pm 0.040$\,\AA, indicating compositional variety without extreme dimensional mismatch. Across the dataset, 85 unique A-B pairings are represented, with Li-V constituting the modal combination. This compositional design enables systematic exploration of size-dependent phenomena alongside chemical effects.

\begin{figure*}[h]
\centering
\includegraphics[width=0.98\textwidth]{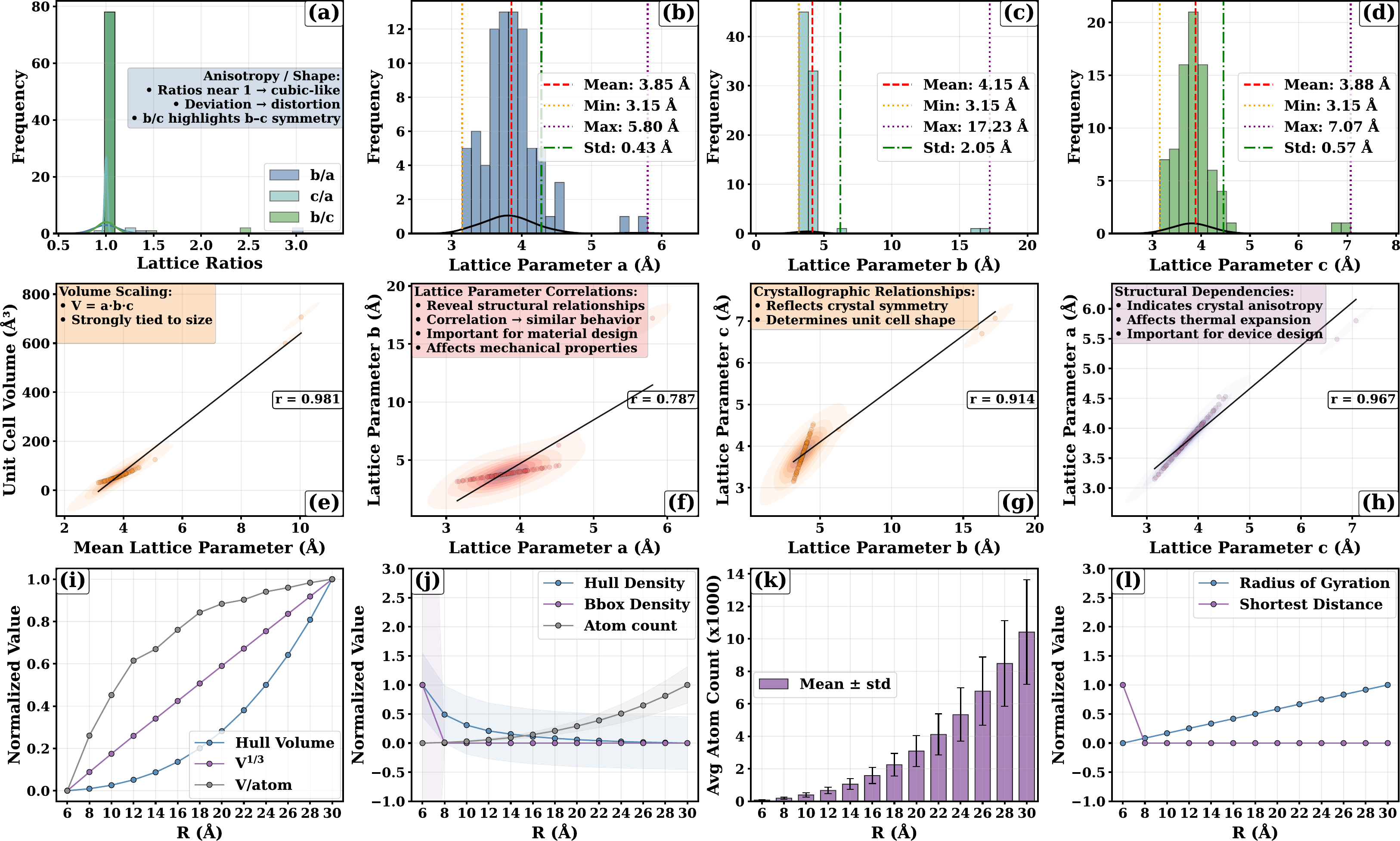}
\caption{Comprehensive analysis of the C2NP dataset. (a--d) Lattice-level statistics of the crystallographic unit cells, including anisotropy ratios and marginal distributions of lattice parameters $a$, $b$, and $c$, highlighting controlled symmetry constraints and directional distortions. (e--h) Unit-cell volume statistics and pairwise correlations between lattice parameters, revealing strong coupling induced by crystallographic structure. (i--l) Radius-dependent nanoparticle properties across $R=6$--30\,\AA, including convex-hull scaling, density and packing behavior, atom counts, and global shape descriptors. Together, these panels demonstrate physically consistent geometric scaling alongside substantial structural variability, establishing C2NP as a rigorous benchmark for both forward nanoparticle generation and inverse lattice recovery.}
\label{fig:dataset_analysis}
\end{figure*}

\subsection{Lattice Geometry and Anisotropy}
\label{subsec:lattice_anisotropy}

Figure~\ref{fig:dataset_analysis}(a) characterizes lattice anisotropy through ratios of lattice parameters. The distributions of $b/a$ and $b/c$ exhibit substantial spread, with means slightly above unity but ranges extending to nearly threefold elongation. This indicates that while many unit cells remain close to cubic or pseudocubic symmetry, a nontrivial fraction display pronounced anisotropy along a single crystallographic direction. In contrast, the $c/a$ ratio is tightly concentrated around $1.0$, with minimal variance, suggesting that distortions preferentially occur along the $b$ axis rather than uniformly across all lattice vectors. This controlled anisotropy provides a balanced setting in which models must handle both near-isotropic and strongly distorted lattices without encountering unrelated crystal topologies.

\subsection{Unit-Cell Parameter Distributions}
\label{subsec:lattice_params}

Figures~\ref{fig:dataset_analysis}(b)--(d) report the marginal distributions of lattice parameters $a$, $b$, and $c$: while $a$ and $c$ are narrowly distributed with modest standard deviations and compact ranges, consistent with approximately cubic metrics, $b$ spans a much wider interval from near-cubic to highly elongated cells and exhibits the largest variance, ensuring that lattice inference is not trivially reducible to isotropic scaling and instead requires models to resolve directional distortions explicitly. Unit-cell volumes (Figure~\ref{fig:dataset_analysis}(e)) span more than an order of magnitude due to the combined effects of lattice elongation and chemical variation, and the large standard deviation relative to the mean shows that volume cannot be reliably inferred from a single lattice constant alone, reinforcing the need for coordinated reasoning across all geometric degrees of freedom.

\subsection{Correlations Between Lattice Parameters}
\label{subsec:lattice_correlations}

Figures~\ref{fig:dataset_analysis}(f)--(h) show strong pairwise correlations among lattice parameters. The tight coupling between $a$ and $c$ ($r=0.967$) reflects symmetry constraints shared across the dataset, while the slightly weaker correlations involving $b$ ($r=0.787$ for $a$--$b$ and $r=0.914$ for $b$--$c$) are consistent with anisotropic distortions concentrated along a single axis. These correlations imply that lattice parameters cannot be predicted independently: accurate reconstruction requires learning joint geometric structure rather than marginal statistics.

\subsection{Radius-Dependent Geometric Scaling}
\label{subsec:radius_scaling}

The bottom row of Figure~\ref{fig:dataset_analysis} summarizes geometric statistics as a function of nanoparticle radius $R$. Convex-hull volume (Figure~\ref{fig:dataset_analysis}(i)) increases monotonically with $R$ and closely follows the expected cubic scaling, while the characteristic length scale $V^{1/3}$ grows approximately linearly with radius. Variance across compositions decreases relative to the mean as $R$ increases, indicating convergence toward bulk-like geometry at larger sizes. Hull volume per atom exhibits a gradual increase with radius, accompanied by growing variance. This trend reflects the diminishing influence of surface truncation at larger $R$, as interior atoms increasingly dominate the structure. At small radii, surface effects introduce substantial heterogeneity, leading to broader distributions.

\subsection{Density and Packing Behavior}
\label{subsec:density_behavior}

Figures~\ref{fig:dataset_analysis}(j) compare atomic densities computed using convex hulls and axis-aligned bounding boxes. Hull-based densities decrease smoothly with increasing radius and stabilize at larger $R$, consistent with the transition from surface-dominated clusters to bulk-like packing. Bounding-box densities show larger variance at small radii, reflecting sensitivity to shape anisotropy and orientation, but converge toward similar values at larger sizes. Together, these measures confirm that C2NP captures realistic size-dependent packing effects without introducing artificial density fluctuations.

\subsection{Atom Count and Structural Extent}
\label{subsec:atom_count}

Figure~\ref{fig:dataset_analysis}(k) reports atom counts as a function of radius. Mean atom counts increase rapidly with $R$, spanning over two orders of magnitude across the dataset. The large variance at each radius reflects compositional and lattice-geometry differences, ensuring that particle size cannot be inferred solely from atom count. This variability further complicates inverse tasks that seek to recover lattice information from finite configurations.

\subsection{Global and Local Length Scales}
\label{subsec:length_scales}

Global size measures, including the radius of gyration (Figure~\ref{fig:dataset_analysis}(l)), scale smoothly with $R$ and exhibit very low variance at larger radii, indicating consistent overall geometry across compositions. In contrast, the shortest interatomic distance remains effectively constant across all radii, with similar mean and variance. This separation of global and local length scales confirms that C2NP preserves chemically meaningful nearest-neighbor distances while varying only the extent of finite-size truncation.


\subsection{Benchmark Tasks}
\label{sec:tasks}

C2NP evaluates scale generalization through two complementary tasks: forward generation of finite nanoparticles from periodic unit cells, and inverse reconstruction of bulk crystallographic parameters from finite atomic configurations. All evaluations are reported separately for in-distribution (interpolative) and out-of-distribution (extrapolative) particle sizes.

\paragraph{Task 1: Unit Cell to Nanoparticle.}

Given a crystallographic unit cell $\mathcal{U}_m$ and a target radius $R$, models must generate a finite nanoparticle $\mathcal{P}$ that preserves the underlying periodic ordering while correctly capturing truncation-induced surface structure. Predicted nanoparticles are compared to reference structures using the following metrics:
 \emph{RMSD}, measuring average atomic displacement after optimal alignment;
 \emph{Hausdorff distance}, capturing worst-case geometric deviation between particle surfaces;
 \emph{Convex-hull volume error}, assessing global shape consistency;
 \emph{Radial distribution function (RDF) error}, comparing local pairwise distance statistics;
\ \emph{Local environment variance}, quantifying heterogeneity in atomic coordination induced by finite-size effects. Exact metric definitions follow standard formulations and are provided in the Supplementary section.

\paragraph{Task 2: Nanoparticle to Unit Cell.}

The inverse task requires recovering bulk crystallographic structure from a finite atomic configuration. Given a nanoparticle $\mathcal{P}$, models predict lattice parameters $\Lambda=(a,b,c,\alpha,\beta,\gamma)$ and the corresponding space-group symmetry $\Gamma$ while performance is evaluated using:
 \emph{Lattice RMSE}, computed over the six lattice parameters;
 \emph{Space-group accuracy}, reporting correct symmetry classification;
 \emph{Joint recovery accuracy}, requiring simultaneous correctness of lattice parameters and space group. This task is intentionally stringent, as surface truncation and broken periodicity obscure global lattice signals.

\paragraph{Metric Rationale.}

The selected metrics jointly capture complementary aspects of scale transfer. RMSD and Hausdorff distance assess atomic-level and worst-case geometric accuracy, while convex-hull volume evaluates global particle shape. RDF error and local environment variance probe preservation of local coordination patterns and surface heterogeneity. For the inverse task, lattice RMSE quantifies geometric reconstruction accuracy, space-group accuracy measures symmetry inference, and joint recovery requires simultaneous success on both. Together, these metrics prevent degenerate solutions and expose failure modes that remain hidden under loss-only evaluation.

\section{Experiments}
\label{sec:exp}

We evaluate a diverse set of state-of-the-art generative models, including CDVAE \cite{xie2021crystal}, DiffCSP \cite{jiao2023crystal}, FlowMM \cite{miller2024flowmm}, MatterGen-MP \cite{zeni2023mattergen}, and ADiT \cite{joshi2025all}, on the C2NP benchmark using the task definitions and size-based splits described in Section~\ref{sec:tasks}. All models are evaluated under identical training and evaluation protocols. Implementation details, hyperparameters, and training configurations are reported in Supplementary~\ref{sec:appenImplement}, Normalized extended experimental results are provided in Supplementary~\ref{sec:appenExperiment}, where all non-time metrics are robustly normalized using a median--MAD z-score followed by a logistic mapping to produce unitless scores, enabling fair comparison across heterogeneous metrics with widely different scales and distributions while preserving relative performance ordering.

\subsection{Task 1: Unit Cell to Nanoparticle Generation}

Table~\ref{tab:merged_task1_task2_results} summarizes performance on the forward nanoparticle generation task, where all non-time metrics are robustly normalized into unitless scores in $(0,1)$ (higher is better). Despite achieving similarly high normalized loss scores (all near $0.61$), ADiT, DiffCSP, FlowMM, and MatterGen fail to produce structurally meaningful nanoparticles: their normalized RMSD, Hausdorff, and convex-hull volume scores cluster between $0.34$ and $0.54$, indicating weak geometric fidelity even when the training objective appears well optimized. In contrast, CDVAE attains a substantially lower normalized loss score ($0.14 \pm 0.48$) but achieves near-optimal geometry across all structural metrics, with normalized RMSD and Hausdorff scores of $1.00$ and similarly strong convex-hull, RDF, and volume-ratio scores. This qualitative gap persists under out-of-distribution evaluation, where all other models degrade sharply while CDVAE maintains stable, near-optimal performance. These results demonstrate that minimizing training loss is a poor proxy for structural fidelity and that C2NP exposes fundamental failure modes in nanoparticle generation under size extrapolation.

\subsection{Task 2: Nanoparticle to Lattice Inference}

Table~\ref{tab:merged_task1_task2_results} also reports results for the inverse lattice-reconstruction task, with all non-time metrics reported as robustly normalized scores in $(0,1)$. Across all evaluated models, no method succeeds at jointly recovering lattice parameters and space-group symmetry from finite nanoparticles. While DiffCSP and ADiT achieve moderate normalized space-group accuracy (approximately $0.61$--$0.66$), their lattice-parameter recovery remains weak, with normalized RMSE scores clustering between $0.34$ and $0.50$, and joint accuracy fixed at $0.50$ for all methods. CDVAE achieves the strongest lattice recovery (normalized score $1.00$) but fails to translate this into improved joint recovery, highlighting a fundamental disconnect between continuous lattice regression and discrete symmetry classification. Performance is largely unchanged under out-of-distribution evaluation, indicating that failures arise from intrinsic limitations in crystallographic inference rather than overfitting to specific size regimes. Together, these findings establish Task~2 as a stringent and unresolved benchmark for recovering global crystallographic structure from finite, surface-perturbed atomic configurations.

\section{Limitations}
\label{sec:limitations}

Despite its systematic construction, C2NP is a controlled benchmark rather than a direct representation of experimentally synthesized nanostructures. Nanoparticles are generated by deterministic carving from ideal periodic supercells and therefore omit thermal fluctuations, defects, surface reconstructions, ligand effects, and environmental influences present in real materials. These choices enable precise and reproducible evaluation of scale transfer but may limit correspondence with experimental robustness. C2NP is also restricted to a structurally consistent subset of crystalline materials and a limited set of space-group symmetries, reducing confounding factors at the cost of broader crystallographic diversity. Evaluation splits focus on particle size and rotational variation, and do not probe other generalization challenges such as compositional extrapolation or defect tolerance, reflecting a deliberate trade-off between clarity and scope.

\section{Conclusion and Future Work}
\label{sec:conc}

C2NP introduces a systematic benchmark for evaluating whether generative models can reason across structural scales, bridging infinite periodic unit cells and finite nanoparticles. By pairing crystallographic structures with size-resolved nanoparticle ensembles and defining bidirectional generation and reconstruction tasks, C2NP enables controlled assessment of scale generalization under both interpolation and extrapolation. Experimental results demonstrate that current state-of-the-art models often achieve low training losses while failing to preserve geometric structure or recover underlying crystallographic order, particularly under distributional shift.

Future work will extend C2NP along several dimensions. Broadening crystallographic coverage to include additional space groups and structural motifs will enable evaluation across a wider range of lattice geometries. Incorporating relaxed surfaces, defects, and temperature-dependent configurations would increase physical realism and better reflect experimental conditions. Together, these extensions position C2NP as a foundation for systematic progress in generative modeling of crystalline matter across length scales.

\bibliographystyle{named}
\bibliography{main}

\appendix




\section{Supplementary}

\subsection{Dataset Split Details}\label{sec:appenSplits}
The dataset consists of 171,234 XYZ files drawn from 81 materials, divided into three distinct subsets for training and evaluation. The training set contains 72,900 files (42.6\%), the in-distribution test set contains 45,198 files (26.4\%), and the out-of-distribution test set contains 53,136 files (31.0\%). On average, this corresponds to 900 files per material in the training set, 558 files per material in the ID test set, and 656 files per material in the OOD test set. This balanced partition ensures that each material contributes consistently across subsets, while still providing meaningful variation in the number of files per material.  

The dataset is additionally partitioned by particle radius, with each radius value defining a distinct geometric regime. Training data spans 15 intermediate radii, providing broad coverage of typical particle sizes. Evaluation is conducted on disjoint radii reserved for testing: six in-distribution radii, and four out-of-distribution radii that probe extreme size regimes. The number of samples per radius varies across splits, averaging approximately 59 configurations per radius in training, 92 in the in-distribution test set, and 163 in the out-of-distribution test set. This stratification exposes models to a wide range of particle sizes during training while reserving structurally distinct radii for controlled interpolation and extrapolation evaluation, enabling a rigorous assessment of scale generalization.

\section{Implementation Overview (Tasks 1 \& 2)}
\label{sec:appenImplement}

All evaluated models were instantiated from their official codebases to preserve the published architectures and training protocols. Experiments were run on a single RTX~4070 GPU. Each model was trained to convergence on the fixed data splits, and reported numbers are averaged across three random seeds.

\subsection{Shared Components Across Tasks 1 \& 2}

We use the following common notation: $E_a$ (atom-embedding size), $C$ (hidden width), $H$ (attention heads), $L$ (depth), $E_t$ (time-embedding size), $E_r$ (radius-embedding size; Task~1 only), $d_{\rm lat}$ (VAE latent size; when applicable), and $N_{\rm SG}$ (number of space-group classes; Task~2 only).

\begin{table}[t]
\centering
\caption{
Shared components across Tasks~1 and~2.
We summarize the common modules and training settings used by all models unless otherwise specified.
}
\label{tab:shared_components}

\small
\setlength{\tabcolsep}{4.6pt}
\renewcommand{\arraystretch}{1.18}

\begin{tabular}{@{}p{0.30\columnwidth} p{0.66\columnwidth}@{}}
\toprule
\textbf{Component} & \textbf{Description} \\
\midrule
Atom embedding
& Learnable atom embedding $\mathrm{Embed}_{\mathrm{atom}}\in\mathbb{R}^{E_a}$. \\

Graph encoder
& SchNet encoder over atomic numbers and 3D positions. \\

Backbone
& Transformer or message-passing backbone with LayerNorm and SiLU feed-forward blocks. \\

Conditioning
& Time embedding ($E_t$); radius embedding ($E_r$) for Task~1. \\

Optimization
& Adam (lr $=10^{-4}$) with learning-rate decay and gradient clipping at $1.0$. \\

Noise modeling
& Diffusion or flow formulation with task-specific $\beta(t)$. \\
\bottomrule
\end{tabular}
\end{table}

\subsection{Task-Specific Heads and Noise Schedules}
Table \ref{tab:task_heads} details the prediction heads and diffusion or flow schedules used for each task. Task 1 focuses on per-atom coordinate prediction and therefore employs noise or velocity heads conditioned on time and particle radius, whereas Task 2 targets global crystallographic inference and uses dedicated heads for lattice regression and space-group classification. Distinct schedule forms are retained when required by individual methods, but all schedules follow standard functional families to maintain comparability.

\begin{table}[t]
\centering
\caption{
Prediction targets and noise schedules used in each task.
}
\label{tab:task_heads}

\small
\setlength{\tabcolsep}{5.0pt}
\renewcommand{\arraystretch}{1.15}

\begin{tabular}{@{}l l l@{}}
\toprule
\textbf{Task} & \textbf{Prediction target} & \textbf{Schedule} \\
\midrule
Task~1 & Atomic coordinates ($\mathbb{R}^{3}$) & Polynomial $\beta(t)$ \\
Task~2 & Lattice parameters + space group & Cubic $\beta(t)$ \\
\bottomrule
\end{tabular}
\end{table}

\subsection{Per-Task, Per-Model Variations}

\subsubsection*{Task 1}

Table \ref{tab:task1_models} reports model-specific architectural choices and hyperparameters for Task 1, which evaluates nanoparticle generation from periodic unit cells. Differences across methods primarily arise in backbone design (Transformer, GCN, or VAE), conditioning strategy, and noise parameterization. All models operate under a shared geometric cutoff and comparable embedding sizes, enabling direct comparison of their ability to generate finite structures under controlled size variation.

\begin{table}[t]
\centering
\caption{
Task~1 baselines (Unit Cell $\rightarrow$ Nanoparticle).
Architectures and the main hyperparameter/schedule choices used in our experiments are reported for reproducibility.
}
\label{tab:task1_models}

\small
\setlength{\tabcolsep}{4.8pt}
\renewcommand{\arraystretch}{1.15}

\begin{tabular}{@{}p{0.22\columnwidth} p{0.40\columnwidth} p{0.34\columnwidth}@{}}
\toprule
\textbf{Model} & \textbf{Architecture} & \textbf{Key settings} \\
\midrule
ADiT
& Diffusion Transformer
& $L{=}1$, $H{=}2$ \\

CDVAE
& Conditional VAE + GCN
& $d_{\mathrm{lat}}{=}4$ \\

DiffCSP
& Diffusion Transformer
& $\beta \in [0.01,2]$ \\

FlowMM
& Flow matching + GCN
& Cosine $\beta(t)$ \\

MatterGen-MP
& Flow matching + MLP
& Linear $\beta(t)$ \\
\bottomrule
\end{tabular}
\end{table}

\subsubsection*{Task 2}

Table \ref{tab:task2_models} summarizes model-specific configurations for Task 2, which requires recovering lattice parameters and space-group symmetry from finite nanoparticles. While all approaches employ a SchNet-based encoder to aggregate atomic information, they differ in how global features are projected, how lattice parameters are constrained, and how symmetry is inferred. Presenting these differences explicitly clarifies the sources of variation in lattice reconstruction performance observed in the experiments.

\begin{table}[t]
\centering
\caption{
Task~2 baselines (Nanoparticle $\rightarrow$ Lattice parameters + Space group).
We list each model family along with its main architectural backbone and the key setting that differs across methods.
}
\label{tab:task2_models}

\small
\setlength{\tabcolsep}{4.8pt}
\renewcommand{\arraystretch}{1.15}

\begin{tabular}{@{}p{0.22\columnwidth} p{0.42\columnwidth} p{0.32\columnwidth}@{}}
\toprule
\textbf{Model} & \textbf{Architecture} & \textbf{Key settings} \\
\midrule
ADiT
& SchNet + Transformer
& Cubic $\beta(t)$ \\

CDVAE
& SchNet + latent inference
& $d_{\mathrm{lat}}{=}4$ \\

DiffCSP
& Diffusion over lattice
& Reverse sampling \\

FlowMM
& Flow matching
& $\beta \in [0.01,2]$ \\

MatterGen-MP
& Diffusion on lattice
& Zero-initialized head \\
\bottomrule
\end{tabular}
\end{table}

\section{Extended Results}\label{sec:appenExperiment}
This section provides a detailed evaluation of state-of-the-art generative models on the C2NP benchmark. The benchmark is structured into distinct tasks that probe different facets of nanoparticle lattice generation, inference, and reconstruction, with metrics carefully designed to capture both structural fidelity and the representation of periodic and surface features. Results are reported separately for in-distribution and out-of-distribution regimes, exposing the strengths and generalization limits of each approach. Evaluation criteria include RMSD, Hausdorff distance, volume and surface agreement, radial distribution function divergence, as well as space group and joint accuracies, together offering a rigorous picture of robustness and failure modes. Tables referenced throughout the section present direct model-to-model comparisons in terms of accuracy and computational efficiency. Future extensions of the benchmark will incorporate additional architectures, with all new results and updates maintained openly in the GitHub repository at \url{https://github.com/KurbanIntelligenceLab/C2NP}.

\subsection{Task 1: Unit Cell $\rightarrow$ Nanoparticle}
\label{sec:appenTask1Extended}

Table~\ref{tab:merged_task1_task2_results} reports performance on Task~1 of C2NP for nanoparticle generation under in-distribution and out-of-distribution size regimes, where all non-time metrics are robustly normalized into unitless scores in $(0,1)$ (higher is better). The results reveal a clear qualitative separation among the evaluated methods. Although ADiT, DiffCSP, FlowMM, and MatterGen achieve similarly high normalized loss scores (all around $0.61$), their geometry-aware performance remains poor: normalized RMSD, Hausdorff, and convex-hull volume scores cluster between $0.34$ and $0.54$, indicating weak structural fidelity despite apparently successful optimization. This demonstrates that minimizing the training loss alone does not guarantee meaningful geometric reconstruction, and that diffusion- and flow-based models can satisfy their objectives while failing to reproduce physically valid nanoparticle structures.

In contrast, CDVAE attains a substantially lower normalized loss score ($0.14 \pm 0.48$) but achieves near-optimal geometry across all structural metrics, with normalized RMSD and Hausdorff scores of $1.00$, convex-hull and RDF scores of $0.86$ and $1.00$, and a perfect volume-ratio score of $1.00$. This indicates that CDVAE’s latent-variable formulation more effectively constrains global structure, prioritizing physically meaningful geometry even when the training objective is not aggressively minimized.

Performance degrades for all methods when moving from in-distribution to out-of-distribution radii, reflecting the increased difficulty of extrapolating across particle sizes. However, CDVAE preserves its relative advantage under OOD evaluation, maintaining near-perfect normalized geometry scores while competing approaches remain clustered below $0.50$ on RMSD, Hausdorff, and hull-volume metrics. This highlights CDVAE’s superior generalization and its ability to maintain physically consistent nanoparticle structures beyond the training regime.

Training efficiency provides additional context. MatterGen is the fastest method, requiring approximately $60$\,s per epoch, followed by FlowMM ($\sim 67$\,s) and DiffCSP ($\sim 75$\,s). ADiT is substantially slower, exceeding $240$\,s per epoch. CDVAE offers a favorable trade-off, training in roughly $71$\,s per epoch while delivering orders-of-magnitude improvements in geometric accuracy. Overall, these results highlight the difficulty of C2NP Task~1: minimizing reconstruction loss alone is insufficient, and evaluation on size-shifted OOD regimes exposes fundamental limitations in current generative models’ ability to capture physically consistent scale transitions.

\begin{table*}[t]
\centering
\caption{Nanoparticle generation from unit cell and lattice/space-group prediction performance. To enable fair comparison across heterogeneous metrics with widely different scales and distributions, we apply a robust normalization to all non-time metrics: values are standardized using a median and median absolute deviation and mapped through a logistic function to produce unitless scores in $(0,1)$ where higher indicates better performance. This reduces sensitivity to outliers and heavy-tailed metrics (e.g., $\Delta$HullVol), prevents artificial saturation at 0/1 as in min--max scaling, and preserves relative performance ordering. Time/Epoch is reported in raw units. Results are averaged across multiple random seeds with standard deviation. Arrows indicate optimization direction: $\downarrow$ = lower is better, $\uparrow$ = higher is better. \textbf{Bold} indicates best; \underline{underlined} indicates second best.}
\label{tab:merged_task1_task2_results}

\begingroup
\normalsize
\setlength{\tabcolsep}{3.0pt}
\renewcommand{\arraystretch}{1.08}
\newcommand{\mstd}[2]{#1\,\ensuremath{\pm}\,#2}
\newcommand{\mbest}[1]{{\bfseries\boldmath #1}}

\resizebox{\textwidth}{!}{%
\begin{tabular}{lccccccc}
\toprule

\multicolumn{8}{c}{\textbf{Nanoparticle Generation, In-Distribution}} \\
\midrule
\textbf{Model} & \textbf{Loss} $\downarrow$ & \textbf{RMSD (\AA)} $\downarrow$ & \textbf{Hausdorff (\AA)} $\downarrow$ & \textbf{$\Delta$ Hull Vol} $\downarrow$ & \textbf{RDF Energy} $\downarrow$ & \textbf{Vol Ratio} $\uparrow$ & \textbf{Time/Epoch (s)} $\downarrow$ \\
\midrule
ADiT & \mbest{\mstd{0.61}{0.00}} & \mstd{0.34}{0.11} & \mstd{0.34}{0.07} & \mstd{0.39}{0.00} & \mstd{0.42}{0.11} & \mstd{0.38}{0.06} & \mstd{242.00}{2.00} \\
CDVAE & \underline{\mstd{0.14}{0.48}} & \mbest{\mstd{1.00}{0.00}} & \mbest{\mstd{1.00}{0.00}} & \mbest{\mstd{0.86}{0.00}} & \mbest{\mstd{1.00}{0.00}} & \mbest{\mstd{1.00}{0.00}} & \mstd{71.00}{4.00} \\
DiffCSP & \mstd{0.61}{0.00} & \underline{\mstd{0.54}{0.25}} & \mstd{0.44}{0.09} & \mstd{0.39}{0.00} & \mstd{0.50}{0.16} & \mstd{0.34}{0.00} & \mstd{75.00}{13.00} \\
FlowMM & \mstd{0.61}{0.01} & \mstd{0.50}{0.00} & \underline{\mstd{0.50}{0.00}} & \mstd{0.39}{0.00} & \mstd{0.34}{0.00} & \underline{\mstd{0.50}{0.00}} & \underline{\mstd{67.00}{1.00}} \\
MatterGen & \mstd{0.61}{0.00} & \mstd{0.00}{0.00} & \mstd{1.00}{0.00} & \underline{\mstd{0.40}{0.00}} & \underline{\mstd{1.00}{0.00}} & \mstd{1.00}{0.00} & \mbest{\mstd{60.00}{1.00}} \\
\midrule

\multicolumn{8}{c}{\textbf{Nanoparticle Generation, Out-of-Distribution}} \\
\midrule
ADiT & \mbest{\mstd{0.61}{0.00}} & \mstd{0.34}{0.06} & \mstd{0.34}{0.00} & \mstd{0.39}{0.00} & \mstd{0.37}{0.05} & \mstd{0.34}{0.00} & - \\
CDVAE & \underline{\mstd{0.14}{0.48}} & \mbest{\mstd{1.00}{0.00}} & \mbest{\mstd{1.00}{0.00}} & \mbest{\mstd{0.86}{0.00}} & \mbest{\mstd{1.00}{0.00}} & \mbest{\mstd{1.00}{0.00}} & - \\
DiffCSP & \mstd{0.61}{0.00} & \underline{\mstd{0.50}{0.08}} & \mstd{0.47}{0.05} & \mstd{0.39}{0.00} & \mstd{0.50}{0.11} & \mstd{0.34}{0.00} & - \\
FlowMM & \mstd{0.61}{0.00} & \mstd{0.50}{0.00} & \underline{\mstd{0.50}{0.00}} & \mstd{0.39}{0.00} & \mstd{0.34}{0.00} & \underline{\mstd{0.50}{0.00}} & - \\
MatterGen & \mstd{0.61}{0.00} & \mstd{0.00}{0.00} & \mstd{1.00}{0.00} & \underline{\mstd{0.40}{0.00}} & \underline{\mstd{0.97}{0.00}} & \mstd{1.00}{0.00} & - \\
\midrule

\multicolumn{8}{c}{\textbf{Lattice-Parameter \& Space-Group Prediction}} \\
\midrule
\textbf{Model} & \multicolumn{2}{c}{\textbf{RMSE (\AA)} $\downarrow$} & \multicolumn{2}{c}{\textbf{SG Accuracy} $\uparrow$} & \multicolumn{2}{c}{\textbf{Joint Accuracy} $\uparrow$} & \textbf{Time/Epoch (s)} $\downarrow$ \\
\cmidrule(lr){2-3}\cmidrule(lr){4-5}\cmidrule(lr){6-7}
& \textbf{ID} & \textbf{OOD} & \textbf{ID} & \textbf{OOD} & \textbf{ID} & \textbf{OOD} & \\
\midrule
ADiT & \mstd{0.34}{0.15} & \mstd{0.34}{0.15} & \underline{\mstd{0.61}{0.00}} & \mbest{\mstd{0.66}{0.23}} & \mbest{\mstd{0.50}{0.00}} & \mbest{\mstd{0.50}{0.00}} & \mstd{83.90}{1.70} \\
CDVAE & \mbest{\mstd{1.00}{0.00}} & \mbest{\mstd{1.00}{0.00}} & \mstd{0.61}{0.00} & \underline{\mstd{0.50}{0.00}} & \mstd{0.50}{0.00} & \mstd{0.50}{0.00} & \mbest{\mstd{54.80}{2.10}} \\
DiffCSP & \underline{\mstd{0.50}{0.17}} & \underline{\mstd{0.50}{0.17}} & \mbest{\mstd{0.61}{0.01}} & \mstd{0.66}{0.53} & \mstd{0.50}{0.00} & \mstd{0.50}{0.00} & \underline{\mstd{55.00}{1.50}} \\
FlowMM & \mstd{1.00}{0.00} & \mstd{1.00}{0.00} & \mstd{0.14}{0.00} & \mstd{0.00}{0.00} & \mstd{0.50}{0.00} & \mstd{0.50}{0.00} & \mstd{79.30}{2.30} \\
MatterGen & \mstd{0.34}{0.15} & \mstd{0.34}{0.15} & \mstd{0.61}{0.00} & \mstd{0.50}{0.00} & \mstd{0.50}{0.00} & \mstd{0.50}{0.00} & \mstd{60.40}{2.10} \\
\bottomrule
\end{tabular}%
}
\endgroup
\end{table*}

\subsection{Task 2: Nanoparticle $\rightarrow$ Lattice}
\label{sec:appenTask2Extended}

Table~\ref{tab:merged_task1_task2_results} also summarizes results for Task~2 of C2NP, which evaluates whether models can recover bulk crystallographic information—both lattice parameters $\Lambda$ and space-group symmetry $\Gamma$—from finite nanoparticle configurations. All non-time metrics are reported as robustly normalized scores in $(0,1)$ (higher is better). The results highlight the intrinsic difficulty of this inverse problem. While several methods achieve moderate space-group accuracy under in-distribution evaluation (e.g., DiffCSP and ADiT with normalized scores around $0.61$), their lattice-parameter recovery remains weak, with normalized RMSE scores clustering between $0.34$ and $0.50$. Even CDVAE, which attains the best normalized lattice performance (score $1.00$), fails to achieve strong joint recovery, with joint accuracy remaining at $0.50$ for all methods, indicating that correct lattice regression and symmetry classification are not being achieved simultaneously.

The out-of-distribution regime further exposes these limitations. Although ADiT and DiffCSP maintain strong normalized space-group accuracy under size extrapolation (up to $0.66$), their lattice-parameter scores remain essentially unchanged, suggesting reliance on memorized correlations rather than generalizable crystallographic reasoning. FlowMM collapses in symmetry prediction, yielding near-zero normalized space-group accuracy, while MatterGen and ADiT show no improvement in lattice recovery under OOD conditions. Across all models, joint accuracy remains flat at $0.50$, demonstrating that success on either the continuous lattice regression or the discrete symmetry classification task does not translate into coordinated recovery of both invariants.

Taken together, these findings underscore the challenge posed by C2NP Task~2. Recovering global crystallographic invariants from finite, surface-perturbed structures requires coordinated reasoning over continuous geometric parameters and discrete symmetry classes, which none of the evaluated architectures successfully achieve. The consistent failure under both ID and OOD settings establishes Task~2 as an open problem and positions C2NP as a stringent diagnostic benchmark for future models designed to reason across structural scales and representation types.

\end{document}